\documentclass[10pt]{article}

\usepackage{graphicx}
\usepackage{amsmath}
\usepackage{amssymb}
\usepackage[cp1251]{inputenc}
\usepackage[T2A]{fontenc}
\numberwithin{equation}{section}
\newcommand{\eqa}{\begin{eqnarray}}
\newcommand{\eeqa}{\end{eqnarray}}
\newcommand{\beq}{\begin{equation}}
\newcommand{\eeq}{\end{equation}}

\begin{document}
\parskip 6pt
\hoffset -1.8cm

\title{Integrable motion of anisotropic space curves and surfaces induced by the Landau-Lifshitz equation}
\author{Zhaidary  Myrzakulova$^{1,2}$\footnote{Email: zrmyrzakulova@gmail.com},  \,           Gulgassyl  Nugmanova$^{1,2}$\footnote{Email: nugmanovagn@gmail.com}, \,
Kuralay Yesmakhanova$^{1,2}$\footnote{Email: krmyrzakulova@gmail.com}\\
and  
Ratbay Myrzakulov$^{1,2}$\footnote{Email: rmyrzakulov@gmail.com}\\
\textsl{$^{1}$Ratbay Myrzakulov Eurasian International Centre for Theoretical Physics}, \\ \textsl{Nur-Sultan, 010009, Kazakhstan}\\   
\textsl{$^{2}$Eurasian National University,
Nur-Sultan, 010008, Kazakhstan}}
\date{}
\maketitle

\begin{abstract} In this paper,  we have
studied the geometrical formulation of  the Landau-Lifshitz equation (LLE) and established its  geometrical  equivalent counterpart as some generalized nonlinear Schr\"{o}dinger equation. When
the anisotropy vanishes, from this result we obtain   the well-known results corresponding  for the isotropic case, i.e. to the Heisenberg ferromagnet equation and the focusing nonlinear Schr\"{o}dinger equation. The relations between the LLE and the differential geometry of space curves in the local and in nonlocal cases are studied. Using the well-known Sym-Tafel formula,  the soliton surfaces  induced by the LLE are  briefly considered. 
\end{abstract}

\tableofcontents

\section{Introduction}

Nonlinear dynamics of one-dimensional anisotropic ferromagnet
can be described by the Landau-Lifshitz equation 
(LLE) \cite{LL}
\begin{equation}
{\bf S}_t={\bf S}\wedge {\bf S}_{xx}+{\bf S}\wedge J{\bf S}, 
\end{equation}
where $J=diag(J_{1}, J_{2}, J_{3})$ and $J_{1}\leq J_{2} \leq J_{3}$, ${\bf S}=(S_{1}, S_{2}, S_{3})$ is the unit spin vector  and ${\bf S}^{2}=S_{3}^{2}+\epsilon S_{1}^{2}+\epsilon S_{2}^{2}=1$. Here $\epsilon=\pm 1$, which gives rise to two different models corresponding to the $su(2)$ ($\epsilon=+1)$ and $su(1,1)$ ($\epsilon=-1)$ cases.  The LLE  is integrable \cite{sklyanin}-\cite{borovik}. It admits several integrable and non-integrable extensions and reductions in 1+1, 2+1 and 3+1 dimensions. The supersymmetric isotropic LLE is one of such generalizations in 1+1 dimensions \cite{RM0}.  In 2+1 dimensions, the LLE  admits the following integrable spin systems:    Ishimori equation \cite{Ishimori}, Myrzakulov-I (M-I) equation \cite{RM1},  Myrzakulov-IX (M-IX) equation \cite{RM00}, Myrzakulov-VIII (M-VIII) equation \cite{Bliev}  and so on. The LLE has  some interesting particular and limiting cases \cite{Faddeev}-\cite{Laire4}. For example, in the case $J_{1}=J_{2}=J_{3}$ that is in the isotropic case, from (1.1) we obtain the following  Heisenberg ferromagnet equation (HFE)    
\begin{equation}
{\bf S}_t={\bf S}\wedge {\bf S}_{xx}. 
\end{equation}
At the same time, in the case $J_{1}=J_{2}$ it is called the LLE with an easy plan. Also it is well-known that the nonlinear Schr\"odinger
equation (NLSE) and sine-Gordon equation (SGE) are some
limiting cases  of LLE under the choice of the spectral
parameter. In this sense,  the LLE
 is a universal integrable model in 1+1 dimensions. The connection of the LLE (as well as the Bloch  equation) with some types of two-level quantum systems was established in \cite{Skrotskii}-\cite{Heller}. 

The gauge equivalence between some integrable equations is  an important tool in
soliton theory \cite{zakharov1}. Gauge equivalence of some integrable systems are well understood both in classical and discrete
versions. In particular, it was shown that the LLE (1.1) is gauge   equivalent to some
nonlinear Schr\"odinger-like equation \cite{Ding1}-\cite{Ma1}.  Gauge equivalence of the LLE  with an easy plan and the nonlinear Schr\"{o}dinger
equation was considered in \cite{Nakamura}-\cite{Huang}. 

There are another type of equivalences between integrable equations  which is  called the geometrical equivalence or Lakshmanan equivalence  \cite{Laksh} (see, also  \cite{RM2}-\cite{RM12} and references therein). The geometrical equivalence is related with the differential geometry of curves and surfaces. The relation between the differential geometry of curves and surfaces and integrable systems were investigated by many authors \cite{RM2}-\cite{57}. These results give  the geometrical formulation of integrable systems. 

Recently, the integrable nonlocal soliton equations were introduced and studied \cite{A1}-\cite{Zhong11}. In particular, the nonlocal HFE were studied \cite{13}-\cite{29}.  Other interesting problems  related with the subjects of this paper are integrable discrete HFE \cite{26}-\cite{61}.

In our  previous works,  we have studied some integrable and nonintegrable continuous classical spin chains. In this paper, we continue our investigations such types of spin chains, namely, we study the LLE. If exactly, in this paper we study the LLE (1.1) from the geometrical point of view. We construct the integrable motion of the anisotropic space curves and anisotropic surfaces induced by the LLE. Also we present the nonlocal versions of the LLE and its equivalent nonlinear Schr\"{o}dinger-like equation. The integrable deformations of  nonlocal space curves and the corresponding Serret-Frenet equation are discussed.

The paper is organized as follows. In Sec. 2, we briefly review some well-known results on the LLE. The geometrical formulation of the LLE in terms of anisotropic space curves is studied in Sec. 3. The nonlocal version of the LLE is presented in Sec. 4. Next, in Sec. 5 we discuss the nonlocal space curves formalism including the corresponding nonlocal Serret-Frenet equation. In Sec. 6, the gauge equivalence between  the LLE  and the  nonlinear Shcr\"{o}dinger - like equation is discussed. In Sec. 7, we briefly consider the anisotropic soliton surfaces by using the famous Sym-Tafel formula. Some generalizations of the LLE are presented in Sec. 8.  The hydrodynamical formulation of the LLE is given in Sec. 9. The quantum Heisenberg model is considered in Sec. 10. The dispersionless LLE is presented in Se. 11 for some particular case. In Sec. 12, the self-similar solutions of the LLE is briefly discussed. The lattice (discrete) version of the LLE is considered in Sec. 13. We give conclusions in the last section.

\section{Landau-Lifshitz equation}
In this section we give some basic informations on the LLE like the equation of motion, three types of Lax pairs, Hamilton formulation, different parametrizations and limiting cases.
\subsection{Equation}
The LLE  has been introduced in \cite{LL} and  describes the dynamics of the orientation of the magnetization (or spin) in ferromagnetic materials. It constitutes nowadays a fundamental tool in the magnetic recording industry, due
to its applications to ferromagnets. The LLE  has the form \cite{Faddeev}
\begin{equation}
{\bf S}_t={\bf S}\wedge {\bf S}_{xx}+{\bf S}\wedge J{\bf S}, 
\end{equation}
where $J=diag(J_{1}, J_{2}, J_{3})$ and $J_{1}\leq J_{2} \leq J_{3}$. In components it takes the form
\begin{eqnarray}
S_{1t}&=&S_{2}S_{3xx}-S_{2xx}S_{3}+S_{2}J_{3}S_{3}-J_{2}S_{2}S_{3},\\ 
S_{2t}&=&S_{3}S_{1xx}-S_{3xx}S_{1}+S_{3}J_{1}S_{1}-J_{3}S_{3}S_{1},\\ 
S_{3t}&=&S_{1}S_{2xx}-S_{1xx}S_{2}+S_{1}J_{2}S_{2}-J_{1}S_{1}S_{2}, 
\end{eqnarray}
or
\begin{eqnarray}
S_{1t}&=&S_{2}S_{3xx}-S_{2xx}S_{3}-J_{23}S_{2}S_{3},\\ 
S_{2t}&=&S_{3}S_{1xx}-S_{3xx}S_{1}-J_{31}S_{3}S_{1},\\ 
S_{3t}&=&S_{1}S_{2xx}-S_{1xx}S_{2}-J_{12}S_{1}S_{2}, 
\end{eqnarray}
where $J_{ij}=J_{i}-J_{j}, \quad I_{ij}=J_{i}+J_{j}$. We also use the following form of the LLE
\begin{eqnarray}
iS_{t}^{+}&=&S^{+}S_{3xx}-S^{+}_{xx}S_{3}+J_{3}S_{3}S^{+}-\frac{1}{2}S_{3}[J_{1}(S^{+}+S^{-})+J_{2}(S^{+}-S^{-})],\\ 
iS^{-}_{t}&=&S^{-}_{xx}S_{3}-S^{-}S_{3xx}-J_{3}S_{3}S^{-}+\frac{1}{2}S_{3}[J_{1}(S^{+}+S^{-})-J_{2}(S^{+}-S^{-})],\\ 
iS_{3t}&=&\frac{1}{2}(S^{-}S^{+}_{xx}-S^{-}_{xx}S^{+})-i(J_{2}-J_{1})S_{1}S_{2}, 
\end{eqnarray}
or
 \begin{eqnarray}
iS_{t}^{+}&=&S^{+}S_{3xx}-S^{+}_{xx}S_{3}+J_{3}S_{3}S^{+}-\frac{1}{2}S_{3}(I_{21}S^{+}-J_{21}S^{-}),\\ 
iS^{-}_{t}&=&S^{-}_{xx}S_{3}-S^{-}S_{3xx}-J_{3}S_{3}S^{-}-\frac{1}{2}S_{3}(J_{21}S^{+}-I_{21}S^{-}),\\ 
iS_{3t}&=&\frac{1}{2}(S^{-}S^{+}_{xx}-S^{-}_{xx}S^{+})-\frac{1}{4}J_{12}(S^{+2}-S^{-2}). 
\end{eqnarray}
The matrix LLE reads as
\begin{equation}
iS_t=\frac{1}{2}[S, S_{xx}+S_{J}], 
\end{equation}
where
\begin{equation}
S=\sum_{k=1}^{3}S_{k}\sigma_{k}=\begin{pmatrix}
S_3 & S^- \\
S^+ & -S_3
\end{pmatrix}, \quad S^2=I, \quad S^\pm=S_1\pm iS_2  
\end{equation}
\begin{equation}
S_{J}=\sum_{k=1}^{3}J_{k}S_{k}\sigma_{k}=\begin{pmatrix}
J_{3}S_3 & J_{1}S_{1}-iJ_{2}S_{2} \\
J_{1}S_{1}+iJ_{2}S_{2} & -J_{3}S_3
\end{pmatrix}.  
\end{equation}
Here $\sigma_{j}$ are Pauli matrices:
\begin{equation}
\sigma_{1}=\begin{pmatrix}
0 & 1 \\
1 & 0
\end{pmatrix}, \quad \sigma_{2}=\begin{pmatrix}
0 & -i \\
i & 0
\end{pmatrix},\quad \sigma_{3}=\begin{pmatrix}
1 &  0\\
0 & -1
\end{pmatrix}.  
\end{equation}
We note that the LLE   is a nonlinear dispersive partial differential equation. The corresponding dispersion relation reads as \cite{Laire1}
\begin{equation}
\omega(k)=\pm \sqrt{|k|^{4}+(J_{21}+J_{23})|k|^{2}+
J_{21}J_{23}},
\end{equation}
what corresponds to  the solutions of the form $e^{i(kx-\omega t)}$ that is to linear sinusoidal waves of frequency $\omega$ and wavenumber $k$. 

\subsection{Lax representation} 
There are several forms of the Lax representations of the LLE: elliptic, $4\times 4$ polynomial and $6\times 6$ polynomial Lax pairs.
\subsubsection{Elliptic Lax pair}
The elliptic Lax representation for the LLE is given by \cite{sklyanin}-\cite{borovik}
\begin{eqnarray}
\Phi_{x}&=&U_{1}\Phi, \\ 
\Phi_{t}&=&V_{1}\Phi, 
\end{eqnarray}
where
\begin{eqnarray}
U_{1}&=&-i\sum_{k=1}^{3}u_{k}S_{k}\sigma_{k},\\ 
V_{1}&=&2i\sum_{k=1}^{3}\frac{u_{1}u_{2}u_{3}}{u_{k}}S_{k}\sigma_{k}-i\sum_{a,b,c=1}^{3}u_{a}\epsilon_{abc}S_{b}S_{cx}\sigma_{a}. 
\end{eqnarray}
Here
\begin{eqnarray}
u_{1}=\frac{\rho}{sn(\lambda,k)}, \quad u_{2}=\frac{\rho dn(\lambda,k)}{sn(\lambda,k)},  \quad u_{3}=\frac{\rho cn(\lambda,k)}{sn(\lambda,k)}  
\end{eqnarray}
and
\begin{eqnarray}
k=\sqrt{\frac{J_{2}-J_{1}}{J_{3}-J_{1}}}, \quad \rho=\frac{1}{2}\sqrt{J_{3}-J_{1}},   \quad 0\leq k\leq 1, \quad \rho\geq 0.  
\end{eqnarray}
Note that  $sn(\lambda,k), \quad cn(\lambda,k), \quad dn(\lambda,k)$ are  Jacobi elliptic functions having the following limit cases
\begin{eqnarray}
sn(\lambda,0)&=&\sin \lambda, \quad  cn(\lambda,0)=\cos\lambda, \quad  dn(\lambda,0)=1, \\
sn(\lambda,1)&=&\tanh\lambda, \quad   cn(\lambda,1)=dn(\lambda,1)=\frac{1}{\cosh\lambda}  
\end{eqnarray}
and
\begin{eqnarray}
u_{a}^{2}-u_{b}^{2}=\frac{1}{4}(J_{b}-J_{a}).  
\end{eqnarray}
\subsubsection{Polynomial $4\times 4$  Lax pair}
The $4\times 4$  polynomial Lax pair  for the LLE has the form \cite{Bordag1}-\cite{Yanovski}
\begin{eqnarray}
U_{1}&=&\frac{1}{2}A_{1}(\lambda+\widetilde{J}),\\ 
V_{1}&=&\frac{1}{2}(\frac{1}{2}A_{1}-[A,A_{1x}]+
\frac{1}{2}A_{2J})(\lambda+\widetilde{J}), 
\end{eqnarray}
where
\begin{eqnarray}
A_{1}=\frac{1}{2}\begin{pmatrix}
0 & S_{1}&S_{2}&S_{3} \\
-S_{1} & 0&S_{3}&-S_{2}\\
-S_{2}&-S_{3}&0&S_{1}\\
-S_{3}&S_{2}&-S_{1}&0
\end{pmatrix},\quad A_{2J}=\frac{1}{2}\begin{pmatrix}
0 & J_{1}S_{1}&J_{2}S_{2}&-J_{3}S_{3} \\
-J_{1}S_{1} & 0&J_{3}S_{3}&J_{2}S_{2}\\
-J_{2}S_{2}&-J_{3}S_{3}&0&-J_{1}S_{1}\\
J_{3}S_{3}&-J_{2}S_{2}&J_{1}S_{1}&0
\end{pmatrix}, \nonumber\\ \widetilde{J}=\begin{pmatrix}
-J_{1}-J_{2}+J_{3} & 0&0&0 \\
0 &-J_{1}+J_{2}-J_{3}&0&0\\
0&0&J_{1}-J_{2}-J_{3}&0\\
0&0&0&J_{1}+J_{2}+J_{3}
\end{pmatrix}.
\end{eqnarray}
\subsubsection{Polynomial $6\times 6$  Lax pair}
The $6\times 6$ polynomial  Lax pair  for the LLE has the form \cite{Bordag1}-\cite{Yanovski}
\begin{eqnarray}
U_{1}&=&\frac{\lambda}{2}\begin{pmatrix}
M({\bf S}) &0 \\
0 & 0
\end{pmatrix}+\frac{1}{2}\begin{pmatrix}
0 & 0 \\
 & -M(J{\bf S})
\end{pmatrix}+\frac{1}{2}ad_{\widetilde{J}}\begin{pmatrix}
M({\bf S}) & 0 \\
0 & 0
\end{pmatrix},\\ 
V_{1}&=&\frac{\lambda^{2}}{4}\begin{pmatrix}
M({\bf S}) &0 \\
0 & 0
\end{pmatrix}+\frac{\lambda}{2}\{\begin{pmatrix}
M({\bf S}\times {\bf S}_{x}) & 0 \\
 & 0
\end{pmatrix}+\frac{1}{2}ad_{\widetilde{J}}\begin{pmatrix}
M({\bf S}) & 0 \\
0 & 0
\end{pmatrix}\}+K, 
\end{eqnarray}
where $ad_{f}g=[f,g]$ and 
\begin{eqnarray}
M({\bf S})=\begin{pmatrix}
0 & S_{3}&-S_{2} \\
-S_{3} & 0&S_{1}\\
S_{2}&-S_{1}&0
\end{pmatrix},\quad \widetilde{J}=\begin{pmatrix}
0 & J \\
J& 0
\end{pmatrix}, \quad  [M({\bf A}), M({\bf B})]=-M({\bf A}\times{\bf B}), \\ K=-\frac{1}{4}\begin{pmatrix}
M(J^{2}{\bf S}) & 0 \\
0&2M(J({\bf S}\times{\bf S}_{x}))
\end{pmatrix}+\frac{1}{4}ad_{J}\begin{pmatrix}
M({\bf S}\times{\bf S}_{x}) & 0 \\
0 &M(J{\bf S})
\end{pmatrix}.
\end{eqnarray}
From the compatibility condition $U_{1t}-V_{1x}+[U_{1},V_{1}]=0$ we obtain 
\begin{eqnarray}
M({\bf S})_{t}+[M({\bf S}), M({\bf S})_{xx}]+[M({\bf S},M(J{\bf S})]=0,
\end{eqnarray}
which is the LLE.
\subsection{Hamiltonian structure}
The  LLE can be written in the Hamiltonian form as \cite{Faddeev}
\begin{eqnarray}
{\bf S}_{t}=\{H,{\bf S}\},  
\end{eqnarray}
where the  Hamilton function is given by
\begin{eqnarray}
H=\frac{1}{2}\int\left({\bf S}_{x}^{2}-J_{1}S_{1}^{2}-J_{2}S_{2}^{2}-J_{3}S_{3}^{2}\right)dx.    
\end{eqnarray}
The  Poisson bracket for the two functionals $A, B$ has the form
\begin{eqnarray}
\{A, B\}=-\sum_{a,b,c=1}^{3}\epsilon_{abc}\int \frac{\delta A}{\delta S_{a}}\frac{\delta B}{\delta S_{b}}S_{c} dxdy.  
\end{eqnarray}
In particular, for the components of the spin vector the corresponding Poisson brackets read as
\begin{eqnarray}
\{S_{a}(x), S_{b}(y)\}=-\epsilon_{abc}S_{c}(x)\delta(x-y).  
\end{eqnarray}
\subsection{Different parametrizations}
The LLE can be written in the different forms that follow from the different parametrizations of the spin vector ${\bf S}$. In this subsection we going to present some of these forms of the LLE. 
\subsubsection{$(\theta-\varphi)$ - form}
Consider the following angle parametrization of the LLE
\begin{eqnarray}
S^{+}=e^{i\varphi}\sin \theta, \quad S^{-}=e^{-i\varphi}\sin \theta, \quad S_{3}=\cos\theta, 
\end{eqnarray}
where $\theta(x,t), \varphi(x,t)$ are  some real functions. 
Then the LLE takes the form
\begin{eqnarray}
\theta_{t}+\varphi_{xx}\sin\theta+2\varphi_{x}\theta_{x}\cos\theta+J_{2}\sin\theta\sin\varphi&=&0,\\
\varphi_{t}\sin\theta-\theta_{xx}+(\varphi_{x}^{2}+J_{3}-J_{1}\cos\varphi)\sin\theta\cos\theta&=&0. 
\end{eqnarray}
\subsubsection{$w$ - form}
Let us introduce the new complex function $w$ as
\begin{eqnarray}
w=\frac{S^{+}}{1+S_{3}}=\tan\frac{\theta}{2}e^{i\varphi}. 
\end{eqnarray}
Then in terms of this function, the LLE takes the form
\begin{eqnarray}
iw_{t}+w_{xx}-\frac{2\bar{w}w_{x}^{2}}{1+|w|^{2}}-\frac{1}{2(1+|w|^{2})}\{[(2J_{3}-I_{21})(1-|w|^{2}+\nonumber\\ J_{12}(w^{2}-\bar{w}^{2})]w+J_{21}(1-|w|^{2})\bar{w}\}=0. 
\end{eqnarray}
\subsubsection{$(p-q)$ form}
We now consider the following parametrization of the spin vector \cite{9909021}
\begin{eqnarray}
S_{1}=p(q^{2}-1)+q, \quad S_{2}=ip(q^{2}+1)+iq, \quad S_{3}=2pq+1. 
\end{eqnarray}
In terms of these new variables $(p-q)$, the LLE reads as
\begin{eqnarray}
p_{t_{1}}&=&-\frac{\partial H}{\partial q}=p_{xx}-2(p^{2}q_{x})_{x},\\
q_{t_{1}}&=&\frac{\partial H}{\partial p}=-q_{xx}-2pq_{x}^{2},
\end{eqnarray}
where 
\begin{eqnarray}
H=p_{x}q_{x}-p^{2}(q_{x}^{2}+r(q))-\frac{1}{12}r^{''}(q)-\frac{1}{2}pr^{'}(q), \quad r^{V}=0, \quad r=-\frac{1}{4}{\bf S}\cdot J{\bf S}, \quad t_{1}=it. 
\end{eqnarray}
\subsubsection{LLE as the Madelung equation}
To study some properties of some nonlinear differential equations  is very useful their   hydrodynamical forms. The same conclusion is correct for the LLE. Let us find its    hydrodynamical form. To derive the  hydrodynamical form of the LLE,  we consider the following Madelung like transformation \cite{Laire4}
\begin{eqnarray}
S^{+}=\sqrt{1-S_{3}^{2}}e^{-i(\phi-0.5\pi)}. 
\end{eqnarray}
Then the LLE (1.1) takes the form
\begin{eqnarray}
\phi_{t}+\left(\frac{S_{3x}}{1-S_{3}^{2}}\right)_{x}-
\frac{S_{3}S_{3x}^{2}}{(1-S_{3}^{2})^{2}}+
\left(\phi_{x}^{2}-J_{23}+
J_{21}\sin^{2}\phi\right)S_{3}&=&0,\\
S_{3t}-\left((1-S_{3}^{2})\phi_{x}\right)_{x}+
0.5J_{21}(1-S_{3}^{2})\sin^{2}2\phi&=&0. 
\end{eqnarray}
This is the Madelung equation form of the LLE (1.1). Note that this Madelung form  will be
essential in the study of solutions of the LLE.
\subsubsection{$\tau_{n}$ - parametrization}
Let us present one of very interesting  parametrizations of the spin vector.  Let  the spin matrix $S$ has the form 
\begin{eqnarray}
S=g^{-1}\sigma_{3}g,
\end{eqnarray}
where $g$ satisfies the following equations
\begin{eqnarray}
g_{x}&=&\begin{pmatrix}
0 & \tau_{n-1}\tau_{n}^{-1} \\
\tau_{n+1}\tau_{n}^{-1} & 0
\end{pmatrix}g,\\
g_{t}&=&i\begin{pmatrix}
\tau_{n-1}\tau_{n+1}\tau_{n}^{-2} & -(\tau_{n-1}\tau_{n}^{-1})_{x} \\
(\tau_{n+1}\tau_{n}^{-1})_{x} & -\tau_{n-1}\tau_{n+1}\tau_{n}^{-2}
\end{pmatrix}g.
\end{eqnarray}
The compatibility condition of this  set of linear equations gives 
\begin{eqnarray}
(D_{t}-D_{x}^{2})\tau_{n+1}\cdot \tau_{n}&=&0, \\
D_{x}^{2}\tau_{n}\cdot\tau_{n}-2\tau_{n+1}\cdot \tau_{n-1}&=&0.
\end{eqnarray}
It is the second member of the TL hierarchy (see e.g. \cite{1006.4600} and references therein). On the other hand, the spin matrix $S$ (2.52) obeys the HFE (as $t\rightarrow it$)
\begin{equation}
2iS_t=[S,S_{xx}]. 
\end{equation} 
At the same time the following functions  
\begin{eqnarray}
r=\tau_{n+1} \tau_{n}^{-1}, \quad q=\tau_{n-1}\tau_{n}
\end{eqnarray}
solves the NLSE (2.61)-(2.62).
\subsection{Limiting cases}
In this subsection we want to present some  well-known particular (limiting) cases of the LLE  (see, e.g. Refs. \cite{Faddeev},\cite{Laire4}).
\subsubsection{HFE}
In the isotropic case $J_{1}=J_{2}=J_{3}$, the LLE turn to the HFE
\begin{equation}
{\bf S}_t={\bf S}\wedge {\bf S}_{xx}. 
\end{equation}
The corresponding Lax pair of the HFE  has  the form
\begin{equation}
U_2=-i\lambda S, \quad V_2=-2i\lambda^2 S+\lambda SS_x. 
\end{equation}
It is well-known  that the HFE (1.2) or (2.59) is gauge \cite{zakharov1} and geometrically \cite{Laksh}  equivalent to the following  NLSE
\begin{eqnarray}
iq_{t}+q_{xx}-2rq^{2}&=&0,\\
ir_{t}-r_{xx}+2qr^{2}&=&0
\end{eqnarray}
with the Lax pair\begin{equation}
U_1=-i\lambda\sigma_3+Q, \quad V_1= -2i\lambda^2\sigma_3+\lambda V_1+V_0. 
\end{equation}
Here 
\begin{equation}
 r=-\epsilon\bar{q}, \quad \sigma_3=\begin{pmatrix}
1 & 0 \\
0 & -1 
\end{pmatrix}, \quad Q=\begin{pmatrix}
0 & q \\
r & 0 
\end{pmatrix}, \quad V_1=2Q, \quad V_0=i\begin{pmatrix}
-rq & q_x \\
-r_x & rq 
\end{pmatrix},
\end{equation}
where $bar$  stands for the complex conjugation and $\epsilon=\pm 1$ signals the focusing $(+)$ and defocusing $(-)$ nonlinearity. Let us introduce the new variables as
\begin{equation}
u=rq, \quad v=-\ln q. 
\end{equation}
Then the NLSE takes the form 
\begin{eqnarray}
iu_{t}-u_{xx}-2(uv_{x})_{x}&=&0,\\
iv_{t}+v_{xx}-v_{x}^{2}+2u&=&0
\end{eqnarray}
which is the Toda lattice  (see e.g. Refs. \cite{9909021}-\cite{1006.4600} and references therein). 
 
\subsubsection{$J_{1}=J_{2}\neq J_{3}$ case}
When $J_{1}=J_{2}\neq J_{3}$, the equation (1.1)   is called the
LLE for a spin chain with an
easy plan. 
\subsubsection{Sine-Gordon equation}
To get the sine-Gordon equation (SGE) from the LLE, we assume that \cite{Faddeev}
\begin{eqnarray}
{\bf S}^{2}&=&R^{2}, \quad S_{1}=-\frac{\beta\pi}{2}, \quad  S_{2}=\sqrt{R^{2}-\frac{\beta^{2}\pi^{2}}{4}}\sin\frac{\varphi\pi}{2}, \quad  S_{3}=\sqrt{R^{2}-\frac{\beta^{2}\pi^{2}}{4}}\cos\frac{\varphi\pi}{2},\\
 J_{2}&=&J_{1}+1, \quad J_{3}=J_{2}+m^{2}R^{-2}=1+J_{1}+m^{2}R^{-2}.
\end{eqnarray}
Then the LLE  takes the form
\begin{equation}
{\bf S}_t=R^{-2}{\bf S}\wedge {\bf S}_{xx}+{\bf S}\wedge J{\bf S}, 
\end{equation}
so that  in the limit $R\rightarrow\infty$, it  turns to the SGE
\begin{eqnarray}
\pi&=&\varphi_{t},\\
\pi_{t}&=&\varphi_{xx}-\frac{m^{2}}{\beta}\sin\beta\varphi
\end{eqnarray}
or
\begin{eqnarray}
\varphi_{tt}-\varphi_{xx}+\frac{m^{2}}{\beta}\sin\beta\varphi=0.
\end{eqnarray}

\subsubsection{NLSE}
Let us now we derive the NLSE from the LLE. To do that, we assume \cite{Faddeev}
\begin{eqnarray}
{\bf S}^{2}=1,\quad S^{+}=\sqrt{2\eta}qe^{2i\kappa \eta^{-1}t},\quad S_{3}=\sqrt{1-2\eta|q|^{2}},\quad J_{2}=J_{1},\quad J_{3}=J_{1}-2\kappa\eta^{-1}, \quad \kappa>0.
\end{eqnarray}
In this case, from the LLE (1.1) as $\eta\rightarrow 0$ we obtain the NLSE
\begin{eqnarray}
iq_{t}+q_{xx}+2\epsilon|q|^{2}q=0.
\end{eqnarray}

\section{Integrable deformations of anisotropic space curves}
In this section, our goal is to present the geometrical formulation of the LLE in terms of space curves. To the end, we now consider the  space curve in the Euclidean space $R^{3}$. Such space curves governed by  the following  Serret-Frenet equation (SFE) and its temporal counterpart equation (see e.g. \cite{Laksh}, \cite{Rog} and references therein)
\begin{equation}
\begin{split}
\begin{pmatrix}
e_1 \\
e_2 \\
e_3
\end{pmatrix}_x
 =C
\begin{pmatrix}
e_1 \\
e_2 \\
e_3
\end{pmatrix}
\end{split},  
\end{equation}
\begin{equation}
\begin{split}
\begin{pmatrix}
e_1 \\
e_2 \\
e_3
\end{pmatrix}_t
 =D
\begin{pmatrix}
e_1 \\
e_2 \\
e_3
\end{pmatrix}
\end{split}, 
\end{equation}
where
\begin{equation}
C=
\begin{pmatrix}
0 & \kappa & \sigma \\
-\kappa & 0 & \tau \\
-\sigma & -\tau & 0
\end{pmatrix}, \quad D=
\begin{pmatrix}
0 & \omega_3 & \omega_2 \\
-\omega_3 & 0 & \omega_1 \\
-\omega_2 & -\omega_1 & 0
\end{pmatrix}. 
\end{equation}
Here the functions $\kappa$  is called normal curvature,   $\sigma$  is called geodesic curvature,  $\tau$ is called geodesic  torsion and
 $\omega_j$ are some real  functions. The later functions  must be expressed in terms of $\kappa, \sigma, \tau$ and their derivatives. The compatibility conditions ${\bf e}_{j x t}= {\bf e}_{j t x}$ of    the equations (3.1)-(3.2) reads as
\begin{equation*}
C_{t}-D_{x}+[C,D]=0.
\end{equation*}
For the elements this equation gives
\begin{equation} \kappa_t=\omega_{3x}-\tau\omega_2+\sigma\omega_1, 
\end{equation}
\begin{equation} \sigma_t=\omega_{2x}-\kappa\omega_1+\tau\omega_3, 
\end{equation}
\begin{equation} \tau_t=\omega_{1x}-\sigma\sigma\omega_3+\kappa\omega_2. 
\end{equation}
Now let us we assume that 
\begin{equation} \kappa=0, \quad  \sigma=r-q, \quad  \tau=-i(r+q), \quad r=-\bar{q}, 
\end{equation}
\begin{equation}
\omega_1=q_{x}-r_{x}-i(q_{21}+q_{12}), \quad \omega_2=-i(r_{x}+q_{x})+q_{21}-q_{12}, \quad \omega_3=-2rq-2iq_{11}. 
\end{equation}
Substituting these expressions of $\kappa, \sigma, \tau, \omega_{j}$  into (3.4)-(3.6), we obtain the following nonlinear Schrodinger-like equation
\begin{eqnarray}
iq_{t}+q_{xx}-2rq^{2}-iR_{12}&=&0,\\
ir_{t}-r_{xx}+2qr^{2}-iR_{21}&=&0,\\
q_{11x}-(qq_{21}-rq_{12})&=&0,
\end{eqnarray}
where 
\begin{eqnarray}
R_{12}=q_{12x}+2qq_{11}, \quad R_{21}=q_{21x}-2rq_{11}.
\end{eqnarray}
It is the case to mention that the system of equations (3.9)-(3.11) is integrable in the sense that it admits the following Lax representation \cite{Ding1}-\cite{Ma1}
\begin{eqnarray}
\Psi_{x}&=&U_{2}\Psi, \\ 
\Psi_{t}&=&V_{2}\Psi, 
\end{eqnarray}
where
\begin{eqnarray}
U_{2}&=&\begin{pmatrix}
0 & q \\
r & 0
\end{pmatrix}, \quad Q=\begin{pmatrix}
q_{11} & q_{12} \\
q_{21} & -q_{11}
\end{pmatrix}, \quad R=\begin{pmatrix}
0 & R_{12} \\
R_{21} & 0
\end{pmatrix}=Q_{x}-[U_{2},Q], \quad R_{21}=-\bar{R}_{12}, \\
V_{2}&=&-i(U_{2}^{2}+U_{2x})\sigma_{3}+Q=
\begin{pmatrix}
-irq+q_{11} & iq_x+q_{12} \\
-ir_x+q_{21} & irq-q_{11}
\end{pmatrix}.
\end{eqnarray}
Let us present here some useful formulas. Let $\Psi$ has the form
\begin{eqnarray}
\Psi=\begin{pmatrix}
g_{1} &-\bar{g}_{2} \\
g_{2} & \bar{g}_{1}
\end{pmatrix}, 
\end{eqnarray}
where $g_{j}$ are some complex functions. From (3.13) follows that
\begin{eqnarray}
[det(\Psi)]_{x}=(|g_{1}|^{2}+|g_{2}|^{2})_{x}=0, 
\end{eqnarray}
that is  $|g_{1}|^{2}+|g_{2}|^{2}=c=const$. Usually we assume $c=1$, so that $|g_{1}|^{2}+|g_{2}|^{2}=1$. Then the function $R$ has the form \cite{Ding1}
\begin{eqnarray}
R=\Psi[(\phi S)_{x}-\frac{i}{2}S_{Jx}]\Psi^{-1}
\end{eqnarray}
or
\begin{eqnarray}
R=\Psi[(\phi S)_{x}-\frac{i}{2}S_{Jx}]\Psi^{-1}=iq[J_{1}(2Im(\bar{g}_{2}\bar{g}_{1}))^{2}+J_{2}(2Re(\bar{g}_{2}\bar{g}_{1}))^{2}+J_{3}(|g_{2}|^{2}-|g_{1}|^{2})]-\nonumber \\ i[J_{1}(g_{1}^{2}+\bar{g}_{2}^{2})Im(\bar{g}_{1}^{2}q-\bar{g}_{2}^{2}\bar{q})+J_{2}(g_{1}^{2}-\bar{g}_{2}^{2})Re(\bar{g}_{1}^{2}q-\bar{g}_{2}^{2}\bar{q})+2J_{3}Re(\bar{g}_{2}g_{1})(\bar{g}_{2}g_{1}\bar{q}+g_{2}\bar{g}_{1}q)],
\end{eqnarray}
where $\phi=J_{1}S_{1}^{2}+J_{2}S_{2}^{2}+J_{3}S_{3}^{2}$ and we used the formula $S_{x}=\Psi^{-1}[\sigma_{3},U_{2}]\Psi$. Now from (3.1)-(3.2), taking into account the last relation, we   obtain the following equation for the unit vector ${\bf e}_{3}$:
\begin{equation}
{\bf e}_{3t}={\bf e}_3\times\left({\bf e}_{3xx}+ J{\bf e}_{3}\right), 
\end{equation}
which is nothing but the LLE (1.1) after the identification ${\bf e}_{3}\equiv{\bf S}$. 

{\bf Theorem 1}. \textit{The Landau-Lifshitz equation (1.1) is geometrically 
equivalent to the nonlinear Schr\"{o}dinger-like equation (3.9)-(3.11). In the isotropic case,   this geometrical 
equivalence is reduced to the well-known geometrical  equivalence
between the HFE and the NLSE \cite{Laksh}}.

\section{Nonlocal Landau-Lifshitz equation}
In this section we want to present the nonlocal version of the LLE. It is well-known that in the nonlocal case, the unit spin vector  ${\bf S}=(S_{1}, S_{2}, S_{3})$ is  the  complex-valued vector \cite{13}. The complex-valued spin vector ${\bf S}$ induced that the unit vectors ${\bf e}_{j}$ become also complex-valued. It means that the curvature $\kappa(t,x)$,  the torsion $\tau(t,x)$ and $\omega_{j}$ are complex-valued functions \cite{Zhong11}. 
Since in the nonlocal case, the spin vector ${\bf S}$  is  the complex-valued vector function, so that we may decompose it  as ${\bf S}={\bf M}+i{\bf L}$, where  ${\bf M}$ and ${\bf L}$ are already real
valued vector functions \cite{13}. These new real vectors  satisfy the following relations
\begin{eqnarray}
{\bf M}^{2}-{\bf L}^{2}=1, \quad {\bf M}\cdot{\bf L}=0.
\end{eqnarray}
From the LLE (1.1) follows that these  real valued
vector functions ${\bf M}$  and ${\bf L}$ satisfy  the following  set of coupled equations:
\begin{eqnarray}
{\bf M}_{t}&=&{\bf M}\wedge{\bf M}_{xx}-{\bf L}\wedge {\bf L}_{xx}+{\bf M}\wedge J{\bf M}-{\bf L}\wedge J{\bf L},\\
{\bf L}_{t}&=&{\bf M}\wedge{\bf L}_{xx}+{\bf L}\wedge{\bf M}_{xx}+{\bf L}\wedge J{\bf M}+{\bf M}\wedge J{\bf L}.
\end{eqnarray}
It is the desired nonlocal LLE. Note that in the isotropic case that is when $J_{1}=J_{2}=J_{3}$ this nonlocal LLE reduces  to the following nonlocal HFE \cite{13}
\begin{eqnarray}
{\bf M}_{t}&=&{\bf M}\wedge{\bf M}_{xx}-{\bf L}\wedge {\bf L}_{xx},\\
{\bf L}_{t}&=&{\bf M}\wedge{\bf L}_{xx}+{\bf L}\wedge{\bf M}_{xx}.
\end{eqnarray}
\section{Nonlocal space curves and Nonlocal Serret-Frenet equation}
As we mentioned in the previous section, in the nonlocal case, the unit vectors ${\bf e}_{j}$ and the functions $\kappa, \tau, \sigma,  \omega_{j}$ are the complex-valied variables. As a consequence of these statements, the corresponding space curves describe by the nonlocal Serret-Frenet equation (SFE). To derive the nonlocal SFE, first we decompose the unit vectors ${\bf e}_{j}$ and functions $\kappa, \sigma, \tau, \omega_{j}$ into the real and image parts as
\begin{eqnarray}
{\bf e}_{1}&=&{\bf m}_{1}+i{\bf l}_{1},\quad {\bf e}_{2}={\bf m}_{2}+i{\bf l}_{2},\quad {\bf e}_{3}={\bf m}_{3}+i{\bf l}_{3},\\ 
\kappa&=&\kappa_{1}+i\kappa_{2}, \quad \sigma=\sigma_{1}+i\sigma_{2}, \quad \tau=\tau_{1}+i\tau_{2}, \\
\omega_{1}&=&\omega_{11}+i\omega_{12}, \quad \omega_{2}=\omega_{21}+i\omega_{22}, \quad \omega_{3}=\omega_{31}+i\omega_{32},
\end{eqnarray}
where ${\bf m}_{j}, {\bf l}_{j}$ are the real valued vector functions and  $\kappa_{j}, \sigma_{j}, \tau_{j}, \omega_{ij}$ are real functions.  From
\begin{eqnarray}
{\bf e}_{j}^{2}=1, \quad {\bf e}_{i}\cdot {\bf e}_{j}=0 (i\neq j),\quad {\bf e}_{i}\wedge{\bf e}_{j}=\epsilon_{ijk}{\bf e}_{k}
\end{eqnarray}
follows that the vectors ${\bf m}_{j}, {\bf l}_{j}$ satisfy the following relations
\begin{eqnarray}
{\bf m}_{j}^{2}-{\bf l}^{2}_{j}=1, \quad {\bf m}_{j}\cdot {\bf l}_{j}=0, \quad {\bf m}_{i}\wedge {\bf m}_{j}-{\bf l}_{i}\wedge {\bf l}_{j}=\epsilon_{ijk}{\bf m}_{k}, \quad {\bf m}_{i}\wedge {\bf l}_{j}+{\bf l}_{i}\wedge {\bf m}_{j}=\epsilon_{ijk}{\bf l}_{k}
\end{eqnarray}
and for the case $i\neq j$ we have
\begin{eqnarray}
{\bf m}_{i}\cdot {\bf m}_{j}-{\bf l}_{i}\cdot {\bf l}_{j}=0,
 \quad {\bf m}_{i}\cdot {\bf l}_{j}+{\bf m}_{j}\cdot {\bf l}_{i}=0.
\end{eqnarray}
We now ready to write the nonlocal SFE which follow from (3.1)-(3.2) and have the forms
\begin{equation}
\begin{pmatrix}
{\bf m}_1 \\
{\bf m}_2 \\
{\bf m}_3
\end{pmatrix}_x
 =C_{1}\begin{pmatrix}
{\bf m}_1 \\
{\bf m}_2 \\
{\bf m}_3
\end{pmatrix}-C_{2}\begin{pmatrix}
{\bf l}_1 \\
{\bf l}_2 \\
{\bf l}_3
\end{pmatrix}, \quad \begin{pmatrix}
{\bf l}_1 \\
{\bf l}_2 \\
{\bf l}_3
\end{pmatrix}_x
 =C_{1}\begin{pmatrix}
{\bf l}_1 \\
{\bf l}_2 \\
{\bf l}_3
\end{pmatrix}+C_{2}\begin{pmatrix}
{\bf m}_1 \\
{\bf m}_2 \\
{\bf m}_3
\end{pmatrix},   
\end{equation}
\begin{equation}
\begin{pmatrix}
{\bf m}_1 \\
{\bf m}_2 \\
{\bf m}_3
\end{pmatrix}_{t}
 =D_{1}\begin{pmatrix}
{\bf m}_1 \\
{\bf m}_2 \\
{\bf m}_3
\end{pmatrix}-D_{2}\begin{pmatrix}
{\bf l}_1 \\
{\bf l}_2 \\
{\bf l}_3
\end{pmatrix}, \quad \begin{pmatrix}
{\bf l}_1 \\
{\bf l}_2 \\
{\bf l}_3
\end{pmatrix}_{t}
 =D_{1}\begin{pmatrix}
{\bf l}_1 \\
{\bf l}_2 \\
{\bf l}_3
\end{pmatrix}+D_{2}\begin{pmatrix}
{\bf m}_1 \\
{\bf m}_2 \\
{\bf m}_3
\end{pmatrix},   
\end{equation}
where $C=C_{1}+iC_{2}, \quad D=D_{1}+iD_{2}$ that is
\begin{equation}
C=\begin{pmatrix}
0 & \kappa_{1} & \sigma_{1} \\
-\kappa_{1} & 0 & \tau_{1} \\
-\sigma_{1} & -\tau_{1} & 0
\end{pmatrix}+i\begin{pmatrix}
0 & \kappa_{2} & \sigma_{2} \\
-\kappa_{2} & 0 & \tau_{2} \\
-\sigma_{2} & -\tau_{2} & 0
\end{pmatrix}, 
 D=\begin{pmatrix}
0 & \omega_{31} & \omega_{21} \\
-\omega_{31} & 0 & \omega_{11} \\
-\omega_{21} & -\omega_{11} & 0
\end{pmatrix}+i\begin{pmatrix}
0 & \omega_{32} & \omega_{22} \\
-\omega_{32} & 0 & \omega_{12} \\
-\omega_{22} & -\omega_{12} & 0
\end{pmatrix}.
\end{equation}
The compatibility conditions ${\bf m}_{jxt}={\bf m}_{jtx}, \quad {\bf l}_{jxt}={\bf l}_{jtx}$ of these equations read as
\begin{equation}
C_{1t}-D_{1x}+[C_{1},D_{1}]-[C_{2},D_{2}]=0, \quad C_{2t}-D_{2x}+[C_{1},D_{2}]+[C_{2},D_{1}]=0.
\end{equation}
In components these equations take the following forms
\begin{eqnarray} \kappa_{1t}&=&\omega_{31x}-(\tau_{1}\omega_{21}-\tau_{2}\omega_{22})+(\sigma_{1}\omega_{11}-\sigma_{2}\omega_{12}),\\
\kappa_{2t}&=&\omega_{32x}-(\tau_{1}\omega_{22}-\tau_{2}\omega_{21})+(\sigma_{2}\omega_{11}+\sigma_{2}\omega_{11}),   \\
\sigma_{1t}&=&\omega_{21x}-(\kappa_{1}\omega_{11}-\kappa_{2}\omega_{12})+(\tau_{1}\omega_{31}-\tau_{2}\omega_{32}),  \\
\sigma_{2t}&=&\omega_{22x}-(\kappa_{1}\omega_{12}-\kappa_{2}\omega_{12})+(\tau_{1}\omega_{32}+\tau_{2}\omega_{31}),  \\
\tau_{1t}&=&\omega_{11x}-(\sigma_{1}\omega_{31}-\sigma_{2}\omega_{32})+(\kappa_{1}\omega_{21}-\kappa_{2}\omega_{22}). \\
\tau_{2t}&=&\omega_{12x}-(\sigma_{1}\omega_{32}+\sigma_{2}\omega_{31})+(\kappa_{1}\omega_{22}+\kappa_{2}\omega_{22}). 
\end{eqnarray}
Now we set 
\begin{equation} \kappa=0, \quad  \sigma=r-q, \quad  \tau=-i(r+q), \quad r=-\bar{q}, 
\end{equation}
\begin{equation}
\omega_1=q_{x}-r_{x}-i(q_{21}+q_{12}), \quad \omega_2=-i(r_{x}+q_{x})+q_{21}-q_{12}, \quad \omega_3=-2rq-2iq_{11}, 
\end{equation}
where $r=\nu\bar{q}(\epsilon_{1}x,\epsilon_{2}t)$.  As in the local case, we can show that the nonlocal  LLE (4.2)-(4.3) is the gauge and geometrical equivalent to the following nonlocal nonlinear Schr\"odinger-like equation
\begin{eqnarray}
iq_{t}+q_{xx}-2rq^{2}-iR_{12}(r,q)&=&0,\\
ir_{t}-r_{xx}+2qr^{2}-iR_{21}(r,q)&=&0,\\
q_{11x}-(qq_{21}-rq_{12})&=&0,
\end{eqnarray}
where 
\begin{eqnarray}
R_{12}(r,q)=q_{12x}+2qq_{11}, \quad R_{21}(r,q)=q_{21x}-2rq_{11}, \quad r=\nu\bar{q}(\epsilon_{1}x,\epsilon_{2}t), \quad \nu=\pm 1, \quad \epsilon_{j}^{2}=1.
\end{eqnarray}

\section{Gauge equivalence between the  LLE and NLS-like equation}
The gauge equivalent counterpart of the LLE was constructed in \cite{Ding1}. For the noncompact case, the LLE (1.1) is called the modified LLE, the corresponding gauge equivalent was studied in \cite{Ma1}. Here let us we consider  the relation between the solutions of the LLE (1.1) and the nonlinear Schr\"odinger-like equation (3.9)-(3.11). Let $S$ is the solution of the LLE (1.1) and $g=\Psi$ is the solution linear equations (3.13)-(3.14). Then we have
\begin{equation}
S=g^{-1}\sigma_{3}g, \quad g=A(S+\sigma_{3})=\begin{pmatrix}
f_{1}(S_{3}+1) & f_{1}S^{-} \\
f_{2}S^{+} & -f_{2}(S_{3}+1)
\end{pmatrix}, \quad A=\begin{pmatrix}
f_{1} & 0 \\
0 & f_{2}
\end{pmatrix}=diag(e^{i\gamma}, e^{-i\gamma}), 
\end{equation}
where $\gamma$ is a real function, $det(g)=1$ and $\Delta=-2(S_{3}+1)$. Since 
\begin{equation}
(S+\sigma_{3})^{-1}=-\frac{1}{\Delta}\begin{pmatrix}
S_{3}+1 & S^{-} \\
S^{+} & -(S_{3}+1)
\end{pmatrix}, \quad \Delta=\det(S+\sigma_{3})=-2(S_{3}+1)
\end{equation}
we obtain
\begin{equation}
(S+\sigma_{3})^{-1}(S+\sigma_{3})=-\frac{1}{\Delta}\begin{pmatrix}
S_{3}+1 & S^{-} \\
S^{+} & -(S_{3}+1)
\end{pmatrix}\begin{pmatrix}
S_{3}+1 & S^{-} \\
S^{+} & -(S_{3}+1)
\end{pmatrix}=I.
\end{equation}
Similarly, we calculate
\begin{equation}
g^{-1}=(S+\sigma_{3})^{-1}A^{-1}=-\frac{1}{\Delta}\begin{pmatrix}
S_{3}+1 & S^{-} \\
S^{+} & -(S_{3}+1)
\end{pmatrix}\begin{pmatrix}
e^{-i\gamma} & 0 \\
0 & e^{i\gamma}
\end{pmatrix}=-\frac{1}{\Delta}(S+\sigma_{3})A^{-1}.
\end{equation}
To find the desired relation we recall that
\begin{equation}
S_{x}=g^{-1}[\sigma_{3},U]g=2g^{-1}\begin{pmatrix}
0 & q \\
-r & 
\end{pmatrix}g.
\end{equation}
Hence we get 
\begin{equation}
\begin{pmatrix}
0 & q \\
-r & 0
\end{pmatrix}= \frac{1}{2}A(S+\sigma_{3})S_{x}(S+\sigma_{3})^{-1}A^{-1}=-\frac{1}{2\Delta}A(S+\sigma_{3})S_{x}(S+\sigma_{3})A^{-1}=-\frac{1}{2\Delta}AEA^{-1}
\end{equation}
and
\begin{equation}
(S+\sigma_{3})S_{x}(S+\sigma_{3})=E=\begin{pmatrix}
0 & e_{12} \\
e_{21} & 0
\end{pmatrix}=
2\begin{pmatrix}
0 & S^{-}S_{3x}-(S_{3}+1)S^{-}_{x} \\
S^{+}S_{3x}-(S_{3}+1)S^{+}_{x} & 0
\end{pmatrix}.
\end{equation}
Hence we obtain
\begin{eqnarray}
\begin{pmatrix}
0 & q \\
-r & 0
\end{pmatrix}=\frac{1}{4(S_{3}+1)}\begin{pmatrix}
0 & e_{12}e^{2i\gamma} \\
e_{21}e^{-2i\gamma} & 0
\end{pmatrix}=\nonumber \\
\frac{1}{2(S_{3}+1)}\begin{pmatrix}
0 & [S^{-}S_{3x}-(S_{3}+1)S^{-}_{x}]e^{2i\gamma}  \\
[S^{+}S_{3x}-(S_{3}+1)S^{+}_{x}]e^{-2i\gamma}  & 0
\end{pmatrix},
\end{eqnarray}
or
\begin{eqnarray}
q&=&\frac{1}{2(S_{3}+1)}e_{12}e^{2i\gamma}=\frac{1}{2(S_{3}+1)}[S^{-}S_{3x}-(S_{3}+1)S^{-}_{x}]e^{2i\gamma}, \\
r&=&-\frac{1}{2(S_{3}+1)}e_{21}e^{-2i\gamma}=-\frac{1}{2(S_{3}+1)}[S^{+}S_{3x}-(S_{3}+1)S^{+}_{x}]e^{-2i\gamma},
\end{eqnarray}
where $r=-\bar{q}$. In terms of $\theta-\varphi$ variables these expressions take the form
\begin{eqnarray}
q&=&\frac{(i\varphi_{x}\sin\theta-\theta_{x})}{2}e^{-i\int\varphi_{x}\cos \theta dx}, \\
r&=&\frac{(i\varphi_{x}\sin\theta+\theta_{x})}{2}e^{i\int\varphi_{x}\cos \theta dx},
\end{eqnarray}
where $\epsilon=\pm 1$. Hence we obtain
\begin{eqnarray}
rq=\frac{\varphi_{x}^{2}\sin^{2}\theta+\theta^{2}_{x}}{4}=\frac{1}{4}{\bf S}_{x}^{2}.
\end{eqnarray}
\section{Anisotropic soliton surfaces induced by the LLE}
In this section, our aim is to present  the anisotropic  soliton surfaces induced by the LLE (1.1). To do that, let us recall that the position vector ${\bf r}=(r_{1}, r_{2}, r_{3})$ of the soliton surface satisfies the certain  two equations. Note that in this approach, the matrix form of the position vector defines by the Sym-Tafel formula
\begin{eqnarray}
r=r_{1}\sigma_{1}+r_{2}\sigma_{2}+r_{3}\sigma_{3}=\begin{pmatrix}
r_3 & r^- \\
r^+ & -r_3
\end{pmatrix}=\Phi^{-1}\Phi_{\lambda}.\nonumber
\end{eqnarray}
 Hence we get the following  two equations:
\begin{eqnarray}
r_{x}&=&\Phi^{-1}U_{\lambda}\Phi, \\
r_{t}&=&\Phi^{-1}V_{\lambda}\Phi,
\end{eqnarray}
where 
\begin{eqnarray}
U_{\lambda}&=&\frac{i\rho}{sn^{2}(\lambda,k)}\left[cn(\lambda,k)dn(\lambda,k)S_{1}\sigma_{1}+cn(\lambda,k)S_{2}\sigma_{2}+dn(\lambda,k)S_{3}\sigma_{3}\right], \\
V_{\lambda}&=&i\left[(u_{2}u_{3})_{\lambda}S_{1}\sigma_{1}+(u_{1}u_{3})_{\lambda}S_{2}\sigma_{2}+
(u_{1}u_{2})_{\lambda}S_{3}\sigma_{3}\right]-i\sum_{a,b,c=1}^{3}u_{a\lambda}\epsilon_{abc}S_{b}S_{cx}\sigma_{a}.
\end{eqnarray}
Here we used the following formulas
\begin{eqnarray}
[sn(\lambda)]_{\lambda}=cn(\lambda,k)dn(\lambda,k), \, [cn(\lambda)]_{\lambda}=-sn(\lambda,k)dn(\lambda,k), \, [dn(\lambda)]_{\lambda}=-k^{2}sn(\lambda,k)cn(\lambda,k)
\end{eqnarray}
and
\begin{eqnarray}
sn^{2}(\lambda)+cn^{2}(\lambda)=1, \quad dn^{2}(\lambda)+k^{2}sn^{2}(\lambda)=1, 
\quad dn^{2}(\lambda)-k^{2}cn^{2}(\lambda)=k^{\prime 2}.
\end{eqnarray}
We now can write the first fundamental form of the desired anisotropic soliton surface
\begin{eqnarray}
I={\bf r}_{x}^{2}dx^{2}+2{\bf r}_{x}\cdot{\bf r}_{t}dxdt+{\bf r}_{t}^{2}dt^{2},
\end{eqnarray}
where
\begin{eqnarray}
{\bf r}_{x}^{2}=\frac{1}{2}tr(r_{x}^{2}), \quad {\bf r}_{x}\cdot {\bf r}_{t}=\frac{1}{2}tr(r_{x}r_{t}), \quad {\bf r}_{t}^{2}=\frac{1}{2}tr(r_{t}^{2}).
\end{eqnarray}
Similarly, we can construct the second fundamental form of the soliton surface corresponding to the LLE. In our case, the well-known  Sym-Tafel formula has the form
\begin{equation}
r=\Phi^{-1}\Phi_{\lambda}= \begin{pmatrix}
r_3 & r^- \\
r^+ & -r_3
\end{pmatrix}.
\end{equation}
Using the following expressions  
\begin{equation}
\Phi_{\lambda}=
\begin{pmatrix}
\phi_{1\lambda} & -\bar{\phi}_{2\lambda} \\
\phi_{2\lambda} & \bar{\phi}_{1\lambda}
\end{pmatrix}, \quad \Phi^{-1}=\frac{1}{\det \Phi}\begin{pmatrix}
\bar{\phi}_{1} & \bar{\phi}_{2} \\
-\phi_{2} & \phi_{1}
\end{pmatrix}, \quad \det\Phi=|\phi_{1}|^2+|\phi_{2}|^2,
\end{equation}
we finally have 
\begin{equation}
r=\frac{1}{\det \Phi}\begin{pmatrix}
\bar{\phi}_{1}\phi_{1\lambda}+\bar{\phi}_{2}\phi_{2\lambda}& -\bar{\phi}_{1}\bar{\phi}_{2\lambda}+\bar{\phi}_{2}\bar{\phi}_{1\lambda} \\
-\phi_{2}\phi_{1\lambda}+\phi_{1}\phi_{2\lambda}& \phi_{2}\bar{\phi}_{2\lambda}+\phi_{1}\bar{\phi}_{1\lambda}
\end{pmatrix}.
\end{equation}
Hence for the components of the position vector ${\bf r}=(r_{1}, r_{2}, r_{3})$ we obtain
\begin{equation}
r^{+}=r_{1}+ ir_{2}=\frac{-\phi_{2}\phi_{1\lambda}+\phi_{1}\phi_{2\lambda}}{\det\Phi}, \quad r^{-}=\frac{-\bar{\phi}_{1}\bar{\phi}_{2\lambda}+\bar{\phi}_{2}\bar{\phi}_{1\lambda} }{\det \Phi}, \quad r_{3}=\frac{\bar{\phi}_{1}\phi_{1\lambda}+\bar{\phi}_{2}\phi_{2\lambda}}{\det \Phi}
\end{equation}
or
\begin{eqnarray}
r_{1}&=&\frac{-\phi_{2}\phi_{1\lambda}+\phi_{1}\phi_{2\lambda}-\bar{\phi}_{1}\bar{\phi}_{2\lambda}+\bar{\phi}_{2}\bar{\phi}_{1\lambda}}{2\det\Phi}, \\
r_{2}&=&\frac{-\phi_{2}\phi_{1\lambda}+\phi_{1}\phi_{2\lambda}+\bar{\phi}_{1}\bar{\phi}_{2\lambda}-\bar{\phi}_{2}\bar{\phi}_{1\lambda}}{2i\det\Phi}, \\
r_{3}&=&\frac{\bar{\phi}_{1}\phi_{1\lambda}+\bar{\phi}_{2}\phi_{2\lambda}}{\det \Phi}.
\end{eqnarray}
Let us now we construct the soliton surface corresponding to the 1-soliton solution of the LLE (1.1) which we presented in the previous section. In this case, the components of the position vector are given by (7.13)-(7.15), where 
\begin{eqnarray}
\phi_{1}&=&c_{1}e^{-i(u_{3}x-2u_{1}u_{2}t)}, \quad \phi_{2} =c_{2}e^{i(u_{3}x-2u_{1}u_{2}t)}, \quad \bar{\phi}_{1}=\bar{c}_{1}e^{-\bar{\chi}}, \quad \bar{\phi}_{2} =\bar{c}_{2}e^{\bar{\chi}}\\
\phi_{1\lambda}&=&-i[u_{3\lambda}x-2(u_{1}u_{2})_{\lambda}t]\phi_{01}, \quad \phi_{2\lambda} =i[u_{3\lambda}x-2(u_{1}u_{2})_{\lambda}t]\phi_{02}.
\end{eqnarray}

Thus in this section we have presented the soliton surface given by the position vector ${\bf r}$ corresponding to the 1-soliton solution of the LLE.

\section{Some generalizations of the LLE}
There are  several integrable and nonintegrable generalizations of the LLE. In this section we will present some of them. 
\subsection{Landau-Lifshitz-Schr\"{o}dinger equation}
 One of  generalizations is the Landau-Lifshitz-Schr\"{o}dinger equation (LLSE) which has the form \cite{Sag}
\begin{eqnarray}
iS_t+\frac{1}{2}[S, S_{xx}+S_{J}]+2Q_{x}+i[S,F_{0}]-[Q,SS_{x}]&=&0, \\ 
Q_{t}-F_{0x}+[Q,F_{0}]&=&0, 
\end{eqnarray} 
where
\begin{equation}
 F_{0}=V_{0}+\begin{pmatrix}
a & b \\
c & -a
\end{pmatrix}, \quad Q=\begin{pmatrix}
0 & q \\
r & 0 
\end{pmatrix}, \quad  V_0=i\begin{pmatrix}
-rq & q_x \\
-r_x & rq 
\end{pmatrix},\quad r=\epsilon\bar{q}.
\end{equation}
In the isotropic case $J_{1}=J_{2}=J_{3}$,  this LLSE  takes the form
\begin{eqnarray}
iS_t+\frac{1}{2}[S, S_{xx}]+2Q_{x}+i[S,F_{0}]-[Q,SS_{x}]&=&0, \\ 
Q_{t}-F_{0x}+[Q,F_{0}]&=&0. 
\end{eqnarray} 
This isotropic LLSE admits the following LR \cite{Sag}
\begin{eqnarray}
Z_{x}&=&U_{3}Z, \\ 
Z_{t}&=&V_{3}Z, 
\end{eqnarray} with the Lax pair
\begin{eqnarray}
U_{3}&=&-i\lambda S+Q, \\
V_{3}&=&-2i\lambda^2 S+\lambda (SS_x +2Q)+F_{0}. 
\end{eqnarray}

\subsection{Myrzakulov-XCIX equation}
Another integrable generalization of the LLE is the Myrzakulov-XCIX (M-XCIX) equation \cite{RM3}-\cite{RM4}. The anisotropic M-XCIX equation  reads as
\begin{eqnarray}
iS_{t}+\frac{1}{2}[S, S_{xx}]-4i\lambda_0S_{x}+\frac{1}{\lambda_0+\omega}[S,W]+\frac{1}{2}[S,S_{J}]&=&0,\\
 iW_{x}+({\lambda_0+\omega}) [S, W]&=&0,
\end{eqnarray}
where $\omega=const$, $S={\Sigma}^{3}_{j=1}S_j(x,y,t)\sigma_j$ is a matrix analogue of the spin vector, $W$ is the  potential with the matrix form $W={\Sigma}^{3}_{j=1}W_j(x,y,t)\sigma_j$.
In the isotropic case $J_{1}=J_{2}=J_{3}$ that is  the  isotropic M-XCIX equation   
\begin{eqnarray}
iS_{t}+\frac{1}{2}[S, S_{xx}]+\frac{1}{\omega}[S,W]&=&0,  \\
 iW_{x}+\omega[S, W]&=&0,
\end{eqnarray}
is   integrable by the IST.  Its   LR can be written in the form \cite{RM3}-\cite{RM4}
 \begin{eqnarray}
\Phi_{x}&=&U_{4}\Phi,\\
\Phi_{t}&=&V_{4}\Phi,
\end{eqnarray}  
where the Lax pair $U_{4}$ and $V_{4}$ have the form  
 \begin{eqnarray}
U_{4}&=&-i\lambda S,  \\ 
V_{4}&=&\lambda^2V_2+\lambda V_{1}+\left(\frac{i}{\lambda+\omega}-\frac{i}{a}\right)W.  
\end{eqnarray} 
Here
\begin{eqnarray}
V_2=-2i S,\quad
V_1=SS_{x}.   
\end{eqnarray}

The M-XCIX equation (8.12)-(8.13) is gauge equivalent to the following SMBE \cite{RM3}-\cite{RM4}
\begin{eqnarray}
iq_{t}+q_{xx}+2\delta |q|^2q-2ip&=&0, \\
p_{x}-2i\omega p -2\eta q&=&0,\\
\eta_{x}+\delta(q^{*} p +p^{*} q)&=&0,
\end{eqnarray}
where $q(x,t), p(x,t)$ are complex functions, $\eta(x,t)$ is a real function, $\omega$ is the  real contant. This system is integrable by the IST. The corresponding  LR is given by
\begin{eqnarray}
\Psi_{x}&=&U_{5}\Psi,\\
\Psi_{t}&=&2\lambda U_{5}\Psi+B\Psi, 
\end{eqnarray}  
where $U_{5}$ and $B$ have the forms 
 \begin{eqnarray}
U_{5}&=&-i\lambda \sigma_3+U_0,\\
B&=&B_0+\frac{i}{\lambda+\omega}B_{-1}. 
\end{eqnarray} 
Here
\begin{eqnarray}
U_0=\begin{pmatrix} 0&q\\-q^{*}& 0\end{pmatrix},\quad B_0=-i\delta |q|^2 \sigma_3+i\begin{pmatrix} 0&q_x\\\delta q^{*}_x& 0\end{pmatrix},\quad 
B_{-1}=\begin{pmatrix} \eta&-p\\-\delta p^{*}& -\eta\end{pmatrix}, 
\end{eqnarray}
where $*$ means a complex conjugate and $\delta=\pm1$, so that  $\delta=+1$ corresponds to the attractive interaction and  $\delta=- 1$ to the repulsive interaction respectively.

\subsection{LLE with the self-consistent variable anisotropy}
Let us consider the variable anisotropy  case that is $J_{j}=\frac{1}{\lambda_0+\omega}K_{j}(x,t)$. Then the LLE (1.1) takes the form
\begin{eqnarray}
{\bf S}_{t}+{\bf S}\wedge {\bf S}_{xx}+{\bf S}\wedge K{\bf S}=0,
\end{eqnarray}
where $K(x,t){\bf S}=(K_{1}(x,t)S_{1}+K_{2}(x,t)S_{2}+K_{3}(x,t)S_{3}$. This equation is integrable if we additionally demand  that the  quantity $K{\bf S}$   to satisfy some additional equation, say, the following equation
\begin{eqnarray}
(K{\bf S})_{x}+\mu{\bf S}\wedge K{\bf S}=0.
\end{eqnarray}
Then the  full set of equations has the form
\begin{eqnarray}
{\bf S}_{t}+{\bf S}\wedge {\bf S}_{xx}+{\bf S}\wedge K{\bf S}&=&0,\\
(K{\bf S})_{x}+\mu{\bf S}\wedge K{\bf S}&=&0.
\end{eqnarray}
\subsection{Integrable LLE with the  external magnetic field}
The previous examples of the generalized LLE were  in the absence of an external magnetic field. We now consider the LLE with the  external magnetic field ${\bf H}$ which can be written as  
\begin{eqnarray}
{\bf S}_{t}+{\bf S}\wedge {\bf S}_{xx}+{\bf S}\wedge J{\bf S}+{\bf S}\wedge {\bf H}&=&0,
\end{eqnarray}
where ${\bf H}=(H_{1}, H_{2}, H_{3})$ is the external magnetic field. We now demand that in this LLE with  the  external magnetic field,  the  vector field ${\bf H}$  is variable and satisfies the  following additional equation
\begin{eqnarray}
{\bf H}_{x}+\mu{\bf S}\wedge {\bf H}=0,
\end{eqnarray}
where $\mu=const$. 
Thus we obtain the following set of the closed equations
\begin{eqnarray}
{\bf S}_{t}+{\bf S}\wedge {\bf S}_{xx}+{\bf S}\wedge J{\bf S}+{\bf S}\wedge {\bf H}&=&0,\\
{\bf H}_{x}+\mu{\bf S}\wedge {\bf H}&=&0,
\end{eqnarray}
which is, in fact,  the LLE with the self-consistent external magnetic field ${\bf H}$. This LLE with the self-consistent external magnetic field  in the isotropic case reads as
\begin{eqnarray}
{\bf S}_{t}+{\bf S}\wedge {\bf S}_{xx}+{\bf S}\wedge {\bf H}&=&0,\\
{\bf H}_{x}+\mu{\bf S}\wedge {\bf H}&=&0.
\end{eqnarray}
This equation  is integrable. 
\subsection{Ishimori equation}
The famous anisotropic Ishimori equation (IE)  has the form \cite{Ishimori}
\begin{eqnarray}
{\bf S}_{t}-{\bf S}\wedge ({\bf S}_{xx}+\alpha^{2} {\bf S}_{yy})-u_{y}{\bf S}_{x}-u_{x}{\bf S}_{y}-{\bf S}\wedge J{\bf S}&=&0,\\
u_{xx}-\alpha^{2}u_{yy}+2\alpha^{2}{\bf S}\cdot ({\bf S}_{x}\wedge {\bf S}_{y})&=&0.
\end{eqnarray}
The isotropic IE is integrable and reads as 
\begin{eqnarray}
{\bf S}_{t}-{\bf S}\wedge ({\bf S}_{xx}+\alpha^{2} {\bf S}_{yy})-u_{y}{\bf S}_{x}-u_{x}{\bf S}_{y}&=&0,\\
u_{xx}-\alpha^{2}u_{yy}+2\alpha^{2}{\bf S}\cdot ({\bf S}_{x}\wedge {\bf S}_{y})&=&0,
\end{eqnarray}
or
\begin{eqnarray}
iw_{t}+w_{xx}-\alpha w_{yy}-\frac{2\bar{w}(w_{x}^{2}-w_{y}^{2})}{1+|w|^{2}}+i\beta(u_{x}w_{y}-u_{y}w_{x})&=&0, \\
u_{xx}+\nu u_{yy}-\frac{ 8 Im(w_{x}w_{y})}{1+|w|^{2}}&=&0.
\end{eqnarray}
\subsection{Myrzakulov-IX equation}
The anisotropic Myrzakulov-IX (M-IX) equation is given by 
\begin{eqnarray}
{\bf S}_{t}-{\bf S}\wedge M_{1}{\bf S}+iA_{2}{\bf S}_{x}+iA_{1}{\bf S}_{y}
-{\bf S}\wedge J{\bf S}&=&0,\\
M_{2}u-2\alpha^{2}{\bf S}\cdot ({\bf S}_{x}\wedge {\bf S}_{y})&=&0,
\end{eqnarray}
where  $M_{j}$ and $A_{j}$ are some operators \cite{RM00}-\cite{Bliev}. The isotropic  M-IX equation is integrable. Note that the isotropic M-IX equation is equivalent to the following equation \cite{zakharov}-\cite{maccari}
\begin{eqnarray}
iq_{t}+M_{1}q+vq&=&0,\\
ip_{t}-M_{1}p-vp&=&0,\\
M_{2}v+2M_{1}(pq)&=&0.
\end{eqnarray}
\subsection{Myrzakulov-VIII equation}
The M-IX equation admits some integrable reductions. For example, as $a=b=-1$ it reduces to the anisotropic Myrzakulov-VIII (M-VIII) equation. The M-VIII  reads as \cite{RM00}-\cite{Bliev}
\begin{eqnarray}
iS_{t}+\frac{1}{2}[S,  S_{xx}]+iuS_{x} +[S,S_{J}]&=&0,\\
u_{x}+u_{y}+\frac{1}{4i}tr(S[S_{x},S_{y}])&=&0.
\end{eqnarray}
Note that the isotropic M-VIII  equation is integrable. Its gauge  equivalent counterpart has the form \cite{zakharov}-\cite{maccari}
\begin{eqnarray}
iq_{t}+q_{xx}+vq&=&0,\\
ip_{t}-p_{xx}-vp&=&0,\\
v_{x}+v_{y}+2(pq)_{y}&=&0.
\end{eqnarray}
\subsection{Myrzakulov-XXXIV equation}
The (1+1)-dimensional anisotropic Myrzakulov-XXXIV (M-XXXIV) equation is given by \cite{RM00}-\cite{Bliev}
\begin{eqnarray}
{\bf S}_{t}-{\bf S}\wedge {\bf S}_{xx}+u{\bf S}_{x}-{\bf S}\wedge J{\bf S}&=&0,\\
u_{t}+u_{x}-\beta({\bf S}_{x}^{2})_{x}&=&0.
\end{eqnarray}
In the isotropic case, the M-XXXIV equation  is integrable and is equivalent to the Yajima-Oikawa equation \cite{Bliev}
\begin{eqnarray}
iq_{t}+q_{xx}-vq&=&0,\\
v_{t}+v_{x}+\delta(|q|^{2})_{x}&=&0.
\end{eqnarray}
\subsection{Myrzakulov-I equation}
The anisotropic Myrzakulov-I (M-I)  equation is given by  \cite{RM1}
\begin{eqnarray}
{\bf S}_{t}-{\bf S}\wedge {\bf S}_{xy}-u{\bf S}_{x}-{\bf S}\wedge J{\bf S}&=&0,\\
u_{x}+{\bf S}\cdot ({\bf S}_{x}\wedge {\bf S}_{y})&=&0.
\end{eqnarray}
In the isotropic case, it  takes the form
\begin{eqnarray}
{\bf S}_{t}-{\bf S}\wedge {\bf S}_{xy}-u{\bf S}_{x}&=&0,\\
u_{x}+{\bf S}\cdot ({\bf S}_{x}\wedge {\bf S}_{y})&=&0.
\end{eqnarray}
or 
\begin{eqnarray}
iw_{t}+w_{xy}-iuw_{x}-\frac{2\bar{w}w_{x}w_{y}}{1+|w|^{2}}&=&0, \\
u_{x}+\frac{ 2i(w_{x}\bar{w}_{y}-\bar{w}_{x}w_{y})}{(1+|w|^{2})^{2}}&=&0.
\end{eqnarray}
The isotropic  M-I equation is integrable by the IST method and equivalent to the following (2+1)-dimensional NLSE
\begin{eqnarray}
iq_{t}+q_{xy}-vq&=&0, \\
v_{x}+2(|q|^{2})_{y}&=&0.
\end{eqnarray}
Note that in 1+1 dimension that is when $y=x$, this NLSE reduces to the usual NLSE
\begin{eqnarray}
iq_{t}+q_{xx}+2|q|^{2}q=0.
\end{eqnarray}

\subsection{Landau-Lifshitz-Gilbert  equation}
 From the physical point of view, one of very important generalizations of the LLE is the so-called Landau-Lifshit-Gilbert  equation (LLGE). It was proposed in 1955 by  T. Gilbert  as  a modification of the LLE (1.1)  to incorporate a damping
term. This LLGE is given by \cite{Laire3}
\begin{eqnarray}
{\bf S}_{t}+\beta{\bf S}\wedge {\bf H}_{eff}+\alpha {\bf S}\wedge{\bf S}\wedge {\bf H}_{eff}=0,
\end{eqnarray}
where the vector function ${\bf H}_{eff}$ is the effective
magnetic field as some kind derivative of the magnetic energy of the material. For more details, we refer to the survey \cite{Laire4} and references therein.
\subsection{Landau-Lifshitz-Bloch equation}
The Landau-Lifshitz-Bloch
 equation (LLBE) is one of important generalizations of the LLE. It describes the magnetization dynamics of magnetic particles at high temperatures without the restriction of a fixed magnetization length and
thus allows for its longitudinal relaxation.  The LLBE is given by (see i.g. \cite{1807.00989})
\begin{eqnarray}
{\bf u}_{t}=c_{1}{\bf u}\wedge {\bf H}_{eff}
+c_{2}{\bf u}\wedge ({\bf u}\wedge {\bf H}_{eff})
+c_{3}\frac{1}{|{\bf u}|^{2}}({\bf u}\cdot {\bf H}_{eff}){\bf u},
\end{eqnarray}
where ${\bf u}$ is the spin polarization, $c_{j}$ are constants, ${\bf H}_{eff}$  is effective field. 
\section{LLE as the Madelung equation. Hydrodynamical formulations}
To study some properties of the LLE, it is very useful its  hydrodynamical form. To derive the  hydrodynamical form of the LLE we consider the following Madelung like transformation \cite{Laire4}
\begin{eqnarray}
S^{+}=\sqrt{1-S_{3}^{2}}e^{-i(\phi-0.5\pi)}. 
\end{eqnarray}
Then the LLE (1.1) takes the form
\begin{eqnarray}
\phi_{t}+\left(\frac{S_{3x}}{1-S_{3}^{2}}\right)_{x}-
\frac{S_{3}S_{3x}^{2}}{(1-S_{3}^{2})^{2}}+
\left(\phi_{x}^{2}-J_{23}+
J_{21}\sin^{2}\phi\right)S_{3}&=&0,\\
S_{3t}-\left((1-S_{3}^{2})\phi_{x}\right)_{x}+
0.5J_{21}(1-S_{3}^{2})\sin^{2}2\phi&=&0. 
\end{eqnarray}
This is the Madelung equation form of the LLE (1.1). Note that this Madelung form  will be
essential in the study of solutions of the LLE. It is interesting to present the Madelung equation for the gauge and geometrical equivalent of the LLE, namely,  for the NLS like equation  (3.9)-(3.11). Let $q=\sqrt{\rho}e^{i\varphi}$. For simplicity,  the equation (3.9)-(3.11) we write in the form
\begin{eqnarray}
iq_{t}+q_{xx}-vq=0,
\end{eqnarray}
where 
\begin{eqnarray}
v=2|q|^{2}+iF_{12}, \quad F_{12}=-R_{12}/q, \quad F_{21}=-R_{21}.
\end{eqnarray}
Then for the equation (3.9)-(3.11), the corresponding Madelung equations takes the form
\begin{eqnarray}
\varphi_{t}+\varphi\varphi_{x}+\left(v-\frac{(\sqrt{\rho})_{xx}}{2\sqrt{\rho}}\right)&=&0, \\
\rho_{t}+(\rho\varphi)_{x}&=&0.
\end{eqnarray}
\section{Quantum Heisenberg model}
It is well-known that the LLE (1.1)  is the continuous classical limit of the quantum Heisenberg XYZ model:
\begin{eqnarray}
\hat{H}=-\frac{1}{2}\sum_{j=1}^{N}\left(J_{1}\hat{S}_{1j}\hat{S}_{1(j+1)}+J_{2}\hat{S}_{2j}\hat{S}_{2(j+1)}+J_{3}\hat{S}_{3j}\hat{S}_{3(j+1)}+h\hat{S}_{3j}\right)
\end{eqnarray}
where $h$ is the external magnetic field, $J_{j}$ are real-valued coupling constants. Three particular cases of this model: i) if $J_{1}=J_{2}=J_{3}$, the model is called quantum Heisenberg XXX model; ii) if $J_{1}=J_{2}\neq J_{3}$, it is the quantum Heisenberg XXZ model; iii) if $J_{1}\neq J_{2}\neq J_{3}$, this model is called the uantum Heisenberg XYZ model. Note that quantum Heisenberg XYZ model is one of  examples of the so-called integrable quantum models. Such integrable quantum models can be solved by the quantum inverse scattering methods and/or by the Bethe ansatz. Finally we present the  well-known result, namely, the spin operators $\hat{S}_{j}$ can be  written in terms of auxiliary oscillators $a_{j}$ and their
conjugates $a_{j}^{+}$ as 
\begin{eqnarray}
\hat{S}^{+}_{j}=a_{j}, \quad \hat{S}=a_{j}^{+}(1-a_{j}^{+}a_{j}), \quad \hat{S}_{3j}=\frac{1}{2}-a_{j}^{+}a_{j}.
\end{eqnarray}

\section{Dispersionless LLE}
Let us present the dispersionless LLE. The vector LLE (1.1)  is equivalent to the only one equation for the complex function $S^{+}(x,t)$. This equation has the form (2.11) that is
 \begin{eqnarray}
iS_{t}^{+}=S^{+}S_{3xx}-S^{+}_{xx}S_{3}+J_{3}S_{3}S^{+}-\frac{1}{2}S_{3}(I_{21}S^{+}-J_{21}S^{-}),
\end{eqnarray}
where $S^{-}=\bar{S}^{+}, \quad S_{3}=\sqrt{1-|S^{+}|^{2}}$. Let us consider the transformation
\begin{eqnarray}
S^{+}=\sqrt{u}e^{i\epsilon^{-1}s},
\end{eqnarray}
where $u=1-S_{3}^{2}$ and $s$ are  some real functions, $\epsilon$ is some real parameter. In our case, the equation (11.1) takes the form
 \begin{eqnarray}
i\epsilon S_{t}^{+}=\epsilon^{2}\left(S^{+}S_{3xx}-S^{+}_{xx}S_{3}\right)+J_{3}S_{3}S^{+}-\frac{1}{2}S_{3}(I_{21}S^{+}-J_{21}S^{-}).
\end{eqnarray}
For simplicity, let us consider the case $J_{1}=J_{2}$ that is the LLE with an easy plan. In this case, the equation (11.3) takes the form
\begin{eqnarray}
i\epsilon S_{t}^{+}=\epsilon^{2}\left(S^{+}S_{3xx}-S^{+}_{xx}S_{3}\right)+J_{3}S_{3}S^{+}-J_{1}S_{3}S^{+}.
\end{eqnarray}
Hence and using the transformation (11.2) we obtain the following set of equations
\begin{eqnarray}
s_{t}+\sqrt{1-u}(s_{x}^{2}+J_{31})&=&0, \\
u_{t}+2\sqrt{1-u}(s_{xx}u+s_{x}u_{x})&=&0,
\end{eqnarray}
or
\begin{eqnarray}
v_{t}+[\sqrt{1-u}(v^{2}+J_{31})]_{x}&=&0, \\
u_{t}+2(uv)_{x}\sqrt{1-u}&=&0,
\end{eqnarray}
where $v=s_{x}$. It is the desired dispersionless LLE (dLLE) for the easy plan case. Note that this dLLE is  integrable. Finally, we note that the dispersionless isotropic LLE reads as
\begin{eqnarray}
v_{t}+[v^{2}\sqrt{1-u}]_{x}&=&0, \\
u_{t}+2(uv)_{x}\sqrt{1-u}&=&0.
\end{eqnarray}

\section{Self-similar solution of the LLE}
In the theory of the nonlinear differential equations, their  self-similar solutions play an important role \cite{86}-\cite{Laire7}. For that reason, in this section we want just briefly mention some basic facts about the self-similar solutions, as example, for  the LLGE (for details see e.g. \cite{Laire7}). Note that the LLGE  contents the LLE as the particular case. In 1+1 dimension, the LLGE we write as \cite{Laire7} 
\begin{eqnarray}
{\bf S}_{t}-\beta{\bf S}\wedge {\bf S}_{xxx}+
\alpha {\bf S}\wedge({\bf S}\wedge {\bf S}_{xx})=0.
\end{eqnarray}
Recall that the solution of the LLGE is called self-similar, if it  satisfies the condition:
\begin{eqnarray}
{\bf S}(x,t)={\bf S}(\lambda x, \lambda^{2}t),
\end{eqnarray} 
i.e the LLGE  is invariant under this  scaling. In \cite{Laire7}  were studied the  following two types self-similar solutions:
\\
i) the expander self-similar solution
\begin{eqnarray}
{\bf S}(x,t)={\bf f}\left(\frac{x}{\sqrt{t-T}}\right),
\end{eqnarray}
ii) the shrinker self-similar solution
\begin{eqnarray}
{\bf S}(x,t)={\bf f}\left(\frac{x}{\sqrt{T-t}}\right).
\end{eqnarray} 

\section{Lattice  LLE}
Almost  all main integrable systems have their  mirror side in the form of the lattice or discrete integrable  equations \cite{Hoffmann}-\cite{95}.  Here we want briefly present some basic facts of the lattice version of the LLE (LLLE). It  has the form (see e.g. \cite{Faddeev} and references therein) 
\begin{eqnarray}
\frac{d\Gamma^{(n)}_{\alpha}}{dt}=\{H_{LLLE}, \Gamma^{(n)}_{\alpha}\},
\end{eqnarray}
where $\Gamma^{(n)}_{\alpha}$  are some dynamical variables and (for details see e.g. \cite{Faddeev})
\begin{eqnarray}
H_{LLLE}=\sum_{n=1}^{N}\log h(\Gamma^{(n)}_{\alpha}, \Gamma^{(n+1)}_{\alpha}). 
\end{eqnarray}
The isotropic LLLE that is the discrete HFE is given by (\cite{Hoffmann}-\cite{Ding7})
\begin{eqnarray}
{\bf S}_{jt}=2\left[\frac{{\bf S}_{j+1}\wedge{\bf S}_{j}}{1+{\bf S}_{j+1}\cdot{\bf S}_{j}}-\frac{{\bf S}_{j}\wedge{\bf S}_{j-1}}{1+{\bf S}_{j}\cdot{\bf S}_{j-1}}\right].
\end{eqnarray} 

\section{Conclusion}

In this paper,  we have
established  that the LLE (1.1) is geometrical  equivalent to the
generalized nonlinear Schr\"{o}dinger like equation (3.9)-(3.11). 
When
the anisotropy vanishes, from this result follows  the well-known results corresponding  for the continuous isotropic spin chain, i.e. the HFE (1.2).  We have studied the relation between the LLE and the differential geometry of space curves in the local and  nonlocal cases. Also the soliton surfaces  induced by the LLE are  briefly considered using the well-known Sym-Tafel formula.

\section{Acknowledgements}  This work was supported  by  the Ministry of Education  and Science of Kazakhstan, Grant AP08856912.

\end{document}